# Reduced absorption of light by metallic intra-cavity contacts: Tamm plasmon based laser mode engineering.

[1,2] M.A. Kaliteevski and [1] A.A. Lazarenko
[1] Saint- Petersburg Academic University 8/3 Khlopina Str, St Petersburg, 194021, Russia
[2] Ioffe Physicotechnical Institute of RAS 26, Polytechnicheskaya,194021 St-Petersburg, Russia

**Abstract**
It was widely accepted that embedding of metallic layers into optoelectronic structures is detrimental to lasing due absorption in metal. However, recently macroscopic optical coherence and lasing was observed in microcavities with intra-cavity single metallic layer. Here we propose the design of the of microcavity-type structure with two intra-cavity metallic layers which could serve as contacts for electrical pumping. The design of optical modes based on utilizing peculiarities of Tamm plasmon provides vanishing absorption due to fixing of the node of electric field of optical mode to metallic layers. Proposed design can be used for fabrication of vertical cavity lasers with intra-cavity metallic contacts.

During the past few decades, the plasmonic properties of metallic nanostructures have received considerable interest in both fundamental and applied fields [1]. Modern plasmonics is based on the physics of surface plasmons – the states of an electromagnetic field localized at the interface between a metal and dielectric – which are analogues of waveguide mode. Recently, a novel type of localized mode of the electromagnetic field (Tamm plasmon) were predicted theoretically [2] and subsequently demonstrated experimentally [3]. Tamm plasmons (TP) are localized at the interface of a specially designed Bragg reflector and a metal and are the analogues to Fabry-Perot cavity modes.

Tamm plasmons are very feasible to make: they can be obtained by depositing the metal film on top of Bragg reflector (which can contain some active media). The TP provides a very simple way of laterally localizing light in the semicoductor structures (microcavity) [4] and are readily fabricated by basic photolithography without the need of etching through micrometer-scale thick multilayer structures. Coupling of an exciton in a quantum dot to such a photonic dot induced by the TP was successfully used as a source of single photons [5]. Despite metallic mirrors being the most commonly used type of light reflectors in normal life, metallic optical components have not found wide use in optoelectronics. The main obstacle preventing the application of metallic mirrors in optoelectronics is the optical loss and heating of the metal due to optical absorption, which leads to a catastrophic degradation of the mirrors and surrounding materials. However, application of metals is particularly necessary for prospective organic diode lasers, where high charge carrier densities need to be injected into organic transport layers. Many organic materials exhibit very large oscillator strengths in their electronic transitions, providing the potential for large optical gain. Moreover, they exhibit large exciton binding energies of several hundred meV [6], which is orders of magnitude larger than for semiconductor materials and gives rise to the stability of excitons at room temperature. This in turn allows exciton coherences at room temperature. As an example, polariton lasing was reported recently in a single crystal organic microcavity at room temperature [7]. Organic optoelectronic devices are specifically sensitive to heating due to high chemical activity, enhanced molecule mobility and low thermal conductivity of organic materials. It was believed metallic components are detrimental to optical coherence and incompatible to lasing [8]. Nevertheless, hybrid metal-organic structures are now the subject of intensive studies. Typically, two different experimental configurations are considered: on the one hand, traditional structures based on surface plasmons which enhance the interactions of light to the organic materials [9, 10] or structures based on Tamm plasmons on the other. In the latter case organic and metallic layers are embedded into a microcavity and confined by distributed Bragg reflectors (DBRs) [11], leading to the formation of such modes. Recently, the observation of lasing in a Tamm structure composed of silver and GaAs at liquid nitrogen temperature has been reported [12]. A peculiarity of TP localized on thin metal films of sub-wavelength thickness embedded into a microcavity is the occurrence of a node of electric field at the metal. Thus, absorption in such structure is substantially reduced, and makes possible the experimental observation of macroscopic optical coherence [13] and lasing [14] in a microcavity with single metallic layer. For an implementation of the efficient electrical pumping of a microcavity, the two metallic layers should be inserted into the structure. This paper is aimed at a development of the design of laser based on microcavity, with two intra-cavity metallic layers, and with low decay of the eigen-mode providing possibility of lasing.

The structure used for modelling is shown in figure 1. Active area made of optically active organic material DCM+Alq3, is sandwiched between two silver layers and surrounded by two $SiO_2/TiO_2$ Bragg reflectors. On each interface between silver layer and Bragg reflectors, TP can be localized. Another Fabry -Perot mode can be localized between two silver layers. When the frequencies the three modes are matched, three hybrid modes appears, the energy splitting of these three modes is defined by a thickness of intra-cavity metallic layers.

Figure 2 show reflection spectra from the structure shown in figure 1. Three dips can be seen in the spectra, corresponding to three hybrid modes, and the widths of these three dips are different. For comparison, spectra from microcavity without metallic layers and from single Bragg reflector covered by silver layer are also shown. The width of each dip of hybrid mode is larger, then those for the mode of microcavity without metal, but smaller then for bare TP mode.

Each mode have different field distribution. The mode of interest, which will potentially support lasing, should possess the following properties: i) minimized overlapping with metallic layers (to reduce absorption); ii) maximized overlapping with active area (to increase amplification of light). We can vary the field profile of eigen-mode by tuning the phases gained by light propagating through various layers (which is linearly proportional to the thickness of non-absorbing layers; for metals dependence is more complicated). To optimize the design of the structure we can analyse complex energies $w$ of eingen modes of the structures. The latter can be done by solving Maxwell equations for light using transfer matrix method and imposing "outgoing wave" boundary conditions [15] to the field of eigen mode:

$$A\begin{pmatrix}1\\-n_f\end{pmatrix} = \hat{M}(\omega)\begin{pmatrix}n_l\\1\end{pmatrix}$$

(1)

where $\hat{M}$ is the transfer matrix through the structure, $n_l$ and $n_f$ are the refractive indices of the semi-infinite media surrounding the structure, and $A$ is a constant. Optimized mode should have lowest value of decay, defined by imaginary part of complex eigen-energy. Figure 3 shows the depenencies of resonant frequency and decay of hybrid mode with localized within an active area of the structure on the distance between metallic layers. It can be seen, that there is optimal length of active area, when the value of the mode decay is minimized. Note, that for hybrid eigen-mode of the structure with metallic elements, decay in increased only by 50% in respect to the mode of microcavity with zero absorption. To explain this phenomenon, one should analyse the profile of such optimized mode shown in figure 1. It can be seen, the electric field of optimized mode have the node at the centre of active area and on metallic layers. Since absorbing region is placed near at the place of a node, absorption is reduced. It should noted, that for two other eigen-mode field is not localized upon based active area.

Since metallic layer in such structure are adjacent to active area, there will be no ohmic resistance of the structure will be reduced, and wall-plug efficiency of electrically pumped laser will be increased. Also, contacts of this type can provide uniform distribution of pumping current over the active area of large lateral size.

In summary, we have demonstrated possibility of lasing from organic microcavity with two metallic intra-cavity layers which can serve as a contacts. It was demonstrated, that eigen-mode in such structures can be engineered in the way, which provide nodes of electric field upon metallic layer, and absorption in such structure is substantially reduced. Proposed design can be used for fabrication of vertical cavity surface emitting laser with metallic contacts adjacent directly to active area.

This work has been supported by RFBR and FP7 IRSES.

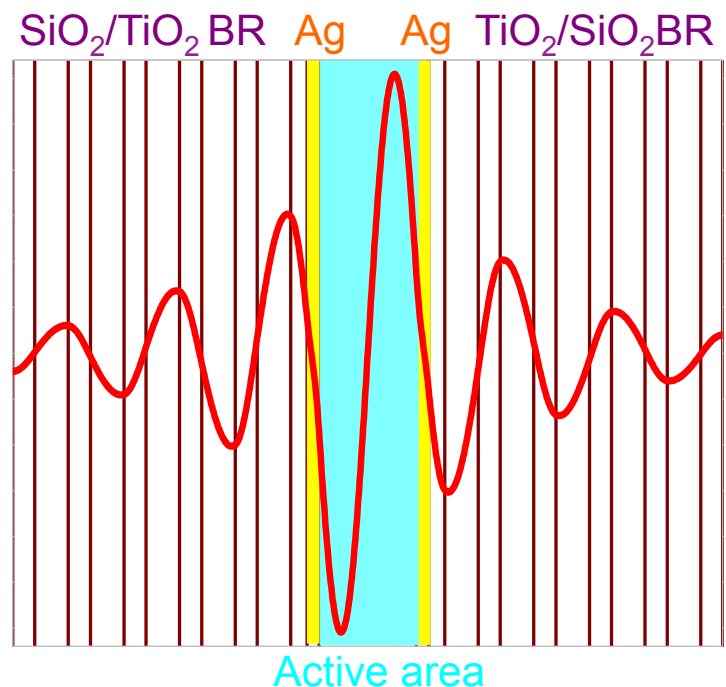

Figure 1 (colour on-line).
Design of the structure: two silver layer of the thickness 40 nm are placed at the boundaries between Bragg mirrors and cavity forming a microcavity structure. Solid line shows the profile of electric field of eigen - modes with minimal decay.

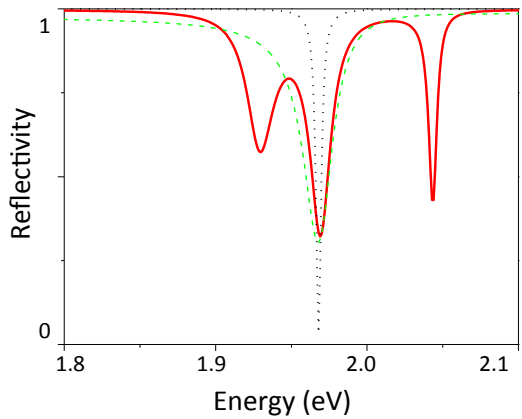

Figure 2 (colour on-line).
Reflection spectra of the structure shown in figure 1 (solid line). For comparison, dotted and dashed lines show the reflection spectra of the microcavity without two metallic intra-cavity layers, and single metallic layer deposited on top of a single Bragg reflector (bare TP mode).

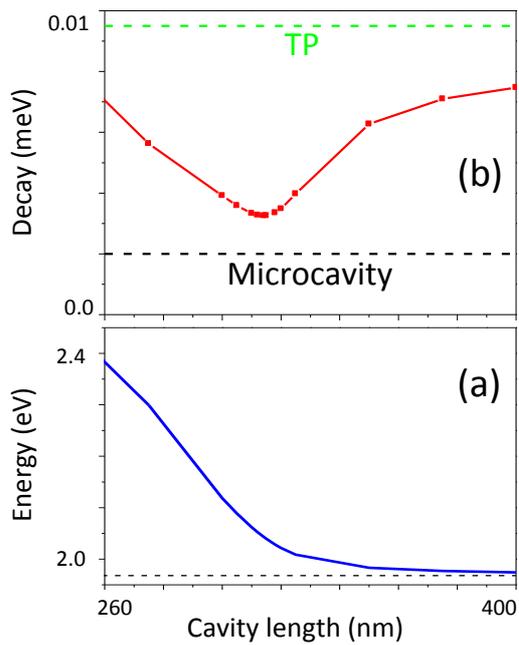

Figure 3 (colour on-line).
Dependence of resonant frequency (a) and decay (b) of the hybrid mode of a microcavity with minimal decay as a function of cavity lentgh. Dashed green and black lines show the decay for TP mode and for the mode of microcavity without metal layers.